# Injecting Uncertainty in Graphs for Identity Obfuscation

Paolo Boldi  Francesco Bonchi  Aristides Gionis  Tamir Tassa

Università degli Studi
Milano, Italy
boldi@dsi.unimi.it

Yahoo! Research
Barcelona, Spain
{bonchi,gionis}@yahoo-inc.com

The Open University
Ra'anana, Israel
tamirta@openu.ac.il

## ABSTRACT

Data collected nowadays by social-networking applications create fascinating opportunities for building novel services, as well as expanding our understanding about social structures and their dynamics. Unfortunately, publishing social-network graphs is considered an ill-advised practice due to privacy concerns. To alleviate this problem, several anonymization methods have been proposed, aiming at reducing the risk of a privacy breach on the published data, while still allowing to analyze them and draw relevant conclusions.

In this paper we introduce a new anonymization approach that is based on injecting *uncertainty* in social graphs and publishing the resulting *uncertain graphs*. While existing approaches obfuscate graph data by adding or removing edges entirely, we propose using a finer-grained perturbation that adds or removes edges *partially*: this way we can achieve the same desired level of obfuscation with smaller changes in the data, thus maintaining higher utility. Our experiments on real-world networks confirm that at the same level of identity obfuscation our method provides higher usefulness than existing randomized methods that publish standard graphs.

## 1. INTRODUCTION

Preserving the anonymity of individuals when publishing social-network data is a challenging problem that has recently attracted a lot of attention [2, 22]. The methods that have been proposed so far for anonymizing social graphs can be classified into three main categories: (1) methods that group vertices into super-vertices of size at least $k$, where $k$ is the required level of anonymity; (2) methods that provide anonymity in the graph via deterministic edge additions or deletions; and (3) methods that add noise to the data in the form of random additions, deletions or switching of edges.

In this paper we introduce a new graph-anonymization method that does not fall in any of the above three categories. Our method injects uncertainty in the existence of the edges of the graph and publishes the resulting *uncertain graph*, that is, a graph where each edge $e$ has an associated probability $\mathbf{p}(e)$ of being present. Injecting a limited amount of uncertainty in the data, in order to reach a desired level of identity obfuscation, is a natural approach [1]. For instance, the $k$-anonymity framework for relational data [25, 28] is typically based on injecting uncertainty by means of attribute generalization; for example, generalizing an exact numerical value to a range of values.

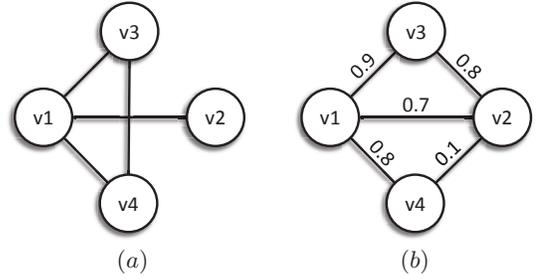

**Figure 1:** (*a*) **A graph;** (*b*) **a possible obfuscation.**

In the context of graph anonymization, our approach can be seen as a generalization of random-perturbation methods, which randomly delete existing edges and add non-existing edges [12]. From a probabilistic perspective, adding a non-existing edge $e$ corresponds to changing its probability $\mathbf{p}(e)$ from 0 to 1, while removing an existing edge corresponds to changing its probability from 1 to 0. In our method, instead of considering only binary edge probabilities, we allow probabilities to take any value in [0, 1], thus allowing for greater flexibility. The underlying intuition is that by using finer-grained perturbation operations, one can achieve the same desired level of obfuscation with smaller changes in the data, thus maintaining higher data utility.

An example of the proposed obfuscation method is shown in Figure 1: The graph (*a*) is the original graph that needs to be obfuscated; the published graph (*b*) is a possible obfuscation. While vertices $v_1$ and $v_2$ are connected in (*a*), in (*b*) they are connected with probability $\mathbf{p}(v_1, v_2) = 0.7$, representing a reduction of 0.3 in the certainty of existence of the edge $(v_1, v_2)$. Vertices $v_3$ and $v_4$, which are connected in (*a*), are no longer connected in the published graph (*b*), i.e., $\mathbf{p}(v_3, v_4) = 0$. Vertices $v_2$ and $v_3$, which were not connected in (*a*), are connected with probability 0.8 in (*b*), corresponding to a partial creation of an edge.

A natural question that arises is how to query and analyze data that is published in the form of an uncertain graph. Hence, in order to prove the practical relevance of our proposal, not only we need to show that the uncertain graph maintains high utility, which we measure as similarity to the original graph in terms of characteristic properties, but also that the computation of these properties can be carried out efficiently. An essential part of our discussion will be devoted to this. Fortunately, an increasing research effort was dedicated in recent years to the topic of querying and mining uncertain graphs [14, 15, 24, 36, 37, 38]: this body of research comes to our aid, providing evidence that useful analysis can be carried out on uncertain graphs.

In this work we achieve the following contributions:

- We introduce and formalize the idea of injecting uncertainty in graphs for identity obfuscation. In particular, we formally define the notion of $(k, \varepsilon)$-*obfuscation* for uncertain graphs (Section 3).

- We provide methods for assessing the level of obfuscation achieved by an uncertain graph with regards to the degree property (Section 4).

- We introduce our method for injecting uncertainty in a graph for $(k, \varepsilon)$-obfuscation (Section 5).

- In Section 6, we discuss several graph statistics and methods to compute them efficiently in uncertain graphs. These statistics are then used in Section 7 to assess the utility of the published uncertain graph.

- Our experimental assessment on three large real-world networks proves that at the same obfuscation levels, our method maintains higher data utility than existing random-perturbation methods.

In the next section we review the relevant literature, while in Section 8 we conclude the paper and suggest future work.

## 2. RELATED WORK

As we already mentioned, methods for anonymizing social networks can be broadly classified into three categories: generalization by means of clustering of vertices; deterministic alteration of the graph by edge additions or deletions; randomized alteration of the graph by addition, deletion or switching of edges.

In the first category, Hay et al. [10, 11] propose to generalize a network by clustering vertices and publishing the number of vertices in each partition together with the densities of edges within and across partitions. Campan and Truta [5] study the case in which vertices contain additional attributes, e.g., demographic information. They propose to cluster the vertices and reveal only the number of intra- and inter-cluster edges. The vertex properties are generalized in such a way that all vertices in the same cluster have the same generalized representation. Tassa and Cohen [29] consider a similar setting and propose a sequential clustering algorithm that issues anonymized graphs with higher utility than those issued by the algorithm of Campan and Truta. Cormode et al. [7, 8] consider a framework where two sets of entities (e.g., patients and drugs) are connected by links (e.g., which patient takes which drugs), and each entity is also described by a set of attributes. The adversary relies upon knowledge of attributes rather than graph structure in devising a matching attack. To prevent matching attacks, their technique masks the mapping between vertices in the graph and real-world entities by clustering the vertices and the corresponding entities into groups. Zheleva and Getoor [33] consider the case where there are multiple types of edges, one of which is sensitive and should be protected. It is assumed that the network is published without the sensitive edges and the adversary predicts sensitive edges based on the observed non-sensitive edges.

In the second category of methods, Liu and Terzi [19] consider the case that a vertex can be identified by its degree. Their algorithms use edge additions and deletions in order to make the graph $k$-*degree anonymous*, meaning that for every vertex there are at least $k - 1$ other vertices with the same degree.

Zhou and Pei [34] consider the case that a vertex can be identified by its radius-one induced subgraph. Adversarial knowledge stronger than the degree is also considered by Thompson and Yao [30], who assume that the adversary knows the degrees of the neighbors, the degrees of the neighbors of the neighbors, and so forth. Zou et al. [35] and Wu et al. [31] assume that the adversary knows the complete graph, and the location of the vertex in the graph; hence, the adversary can always identify a vertex in any copy of the graph, unless the graph has other vertices that are automorphically-equivalent.

In the last category of methods, Hay et al. [12] study the effectiveness of random perturbations for identity obfuscation. They concentrate on degree-based re-identification of vertices. Given a vertex $v$ in the real network, they quantify the level of anonymity that is provided for $v$ by the perturbed graph as $(\max_u \{\Pr(v \mid u)\})^{-1}$, where the maximum is taken over all vertices $u$ in the released graph and $\Pr(v \mid u)$ stands for the belief probability that $u$ is the image of the target vertex $v$. By performing experimentation on the Enron dataset, using various values for the number $h$ of added and removed edges, they conclude that in order to achieve a meaningful level of anonymity for the vertices in the graph, $h$ has to be tuned so high that the resulting features of the perturbed graph no longer reflect those of the original graph.

Ying et al. [32] compare random-perturbation methods to the method of $k$-degree anonymity [19]. They too use the a-posteriori belief probabilities to quantify the level of anonymity. Based on experimentation on two modestly-sized datasets (Enron and Polblogs) they conclude that the deterministic approach for $k$-degree anonymity preserves the graph structure better than random-perturbation methods.

In a more recent study, Bonchi et al. [4] take a different approach, by considering the entropy of the a-posteriori belief probability distributions as a measure of identity obfuscation. The rationale is that while using the a-posteriori belief probabilities is a local measure, the entropy is a global measure that examines the entire distribution of these belief probabilities. Bonchi et al. show that the entropy measure is more accurate than the a-posteriori belief probability, in the sense that the former distinguishes between situations that the latter perceives as equivalent. Moreover, the obfuscation level quantified by means of the entropy is always greater than the one based on a-posteriori belief probabilities. Finally, by means of a thorough experimentation on three large datasets, using several graph statistics and comparing also to Liu and Terzi [19], they demonstrate that random perturbation could be used to achieve meaningful levels of obfuscation while preserving most of the features of the original graph.

## 3. OBFUSCATION BY UNCERTAINTY

Let $G = (V, E)$ be an undirected graph, where $V$ is the set of vertices and $E$ is the set of edges. We write $V_2$ to denote the set of all $\binom{n}{2}$ unordered pairs of vertices from $V$, that is, $V_2 = \{(v_i, v_j) \mid 1 \leq i < j \leq n\}$. The goal is to anonymize the graph $G$ so that the identity of its vertices is obfuscated. We propose to publish $G$ as an uncertain graph $\tilde{G} = (V, \mathbf{p})$, formally defined as follows.

DEFINITION 1. *Given a graph $G = (V, E)$, an uncertain graph on the vertices of $G$ is a pair $\tilde{G} = (V, \mathbf{p})$, where $\mathbf{p} : V_2 \to [0, 1]$ is a function that assigns probabilities to unordered pairs of vertices.*

The original graph $G$ and the uncertain graph $\tilde{G}$ have the same set of vertices $V$. For the sake of clarity, we write $v \in G$ when we speak about a vertex in $G$, and $v \in \tilde{G}$ when we speak about a vertex in $\tilde{G}$.

Since the mere description of an uncertain graph consists of $|V_2| = n(n-1)/2$ probability values, we propose to inject uncertainty only to a small subset of pairs of vertices. Namely, given a graph $G$, we create a subset $E_C \subseteq V_2$ of candidate edges, and then we inject uncertainty only to the pairs of vertices in $E_C$, while we implicitly assume that $\mathbf{p}(u, v) = 0$ for all $(u, v) \notin E_C$. The size of $E_C$ will be set so that $|E_C| = c|E|$, for a small constant $c > 1$. In Section 5 we describe a strategy for selecting $E_C$, given $G$ and a user-defined parameter $c$.

The uncertain graph $\tilde{G}$ induces a collection of *possible worlds* $\mathcal{W}(\tilde{G})$. A possible world $W \in \mathcal{W}(\tilde{G})$ is a graph $W = (V, E_W)$, where $E_W \subseteq E_C$. The edge probabilities in the uncertain graph $\tilde{G}$ imply that the probability of $W$ is

$$\Pr(W) = \prod_{e \in E_W} \mathbf{p}(e) \cdot \prod_{e \in E_C \setminus E_W} (1 - \mathbf{p}(e)). \quad (1)$$

Let us consider the knowledge that an adversary may extract from such an uncertain graph about a given target vertex in $G$. Following the literature, we assume that the adversary knows some vertex property $P$ of his target vertex [4, 12, 19, 30, 31, 32, 34, 35]. Examples of such properties, as discussed in Section 2, are the degree, the degrees of the vertex and its neighbors, and the neighborhood subgraph induced by the target vertex and its neighbors.

Let $\Omega_P$ be the domain in which $P$ takes values, e.g., if $P$ is the degree property then $\Omega_P = \{0, \ldots, n-1\}$. Given an uncertain graph $\tilde{G}$ and a property $P$, for each $v \in \tilde{G}$ and $\omega \in \Omega_P$ we define the probability $X_v(\omega)$ that $v$ originated from a vertex in $G$ with property value $\omega$. Specifically,

$$X_v(\omega) = \sum_{W \in \mathcal{W}(\tilde{G})} \Pr(W) \cdot \chi_{v,\omega}(W), \quad (2)$$

where $\Pr(W)$ is given in Equation (1), and $\chi_{v,\omega}(W)$ is a 0–1 variable that indicates if the vertex $v$ has the property value $\omega$ in the possible world $W$. In other words, $X_v(\omega)$ is the sum of probabilities of all possible worlds in which the vertex $v$ has the given property value $\omega$.

The probabilities $X_v(\omega)$ may be arranged in a $n \times |\Omega_P|$ matrix, where each row corresponds to one vertex $v \in \tilde{G}$ and it gives the corresponding probability distribution $X_v(\omega)$ over all possible values $\omega \in \Omega_P$. The columns of that matrix are proportional to the probability distributions that correspond to property values. More precisely, the normalized column corresponding to property $\omega \in \Omega_P$, i.e.,

$$Y_\omega(v) := \frac{X_v(\omega)}{\sum_{u \in \tilde{G}} X_u(\omega)} \quad (3)$$

is the probability that $v$ is the image in $\tilde{G}$ of a vertex that had the property $\omega$ in $G$.

| $X_v(\omega)$ | deg=0 | deg=1 | deg=2 | deg=3 |
|---|---|---|---|---|
| $v_1$: | 0.006 | 0.092 | 0.398 | 0.504 |
| $v_2$: | 0.054 | 0.348 | 0.542 | 0.056 |
| $v_3$: | 0.020 | 0.260 | 0.720 | 0.000 |
| $v_4$: | 0.180 | 0.740 | 0.080 | 0.000 |

| $Y_\omega(v)$ | deg=0 | deg=1 | deg=2 | deg=3 |
|---|---|---|---|---|
| $v_1$: | 0.023 | 0.064 | 0.229 | 0.900 |
| $v_2$: | 0.208 | 0.242 | 0.311 | 0.100 |
| $v_3$: | 0.077 | 0.180 | 0.414 | 0.000 |
| $v_4$: | 0.692 | 0.514 | 0.046 | 0.000 |

**Table 1: The matrices $X_v(\omega)$ and $Y_\omega(v)$ for the uncertain graph in Figure 1(b) and the degree property.**

EXAMPLE 1. *Consider the uncertain graph in Figure 1(b) and assume property $P_1$. Table 1 gives the corresponding matrix $X_v(\omega)$, in which each row gives the probability distribution regarding the degree of the corresponding vertex in $G$. For instance, the probability that $v_1$ has degree 2 is $0.7 \cdot 0.9 \cdot (1 - 0.8) + 0.7 \cdot (1 - 0.9) \cdot 0.8 + (1 - 0.7) \cdot 0.8 \cdot 0.7 = 0.398$.*

*The columns of $X_v(\omega)$, after normalizing them, give the corresponding $Y_\omega(v)$ distributions for each value of the degree (shown also in Table 1). For instance, if we look for a vertex that has degree 3 in $G$, it is either $v_1$, with probability 0.9, or $v_2$, with probability 0.1.*

To further stress the difference between the two probability distributions, $X_v(\omega)$ and $Y_\omega(v)$, let us consider an uncertain graph $\tilde{G}$ in which all edge probabilities are either 0 or 1 (i.e., a certain graph). Let $\omega$ be some property value in $\Omega_P$ and assume that $P^{-1}(\omega) = \{v_{i_1}, \ldots, v_{i_k}\}$ (namely, there are exactly $k$ vertices with the property $\omega$ in the graph). Then, for all $v \in P^{-1}(\omega)$, $X_v(\omega) = 1$ (since each of them has the property $\omega$ with certainty) and $X_v(\omega') = 0$ for any other property $\omega' \neq \omega$ (since any vertex can have in any certain graph just one property). Furthermore, $X_v(\omega) = 0$ for all $v \notin P^{-1}(\omega)$. Let us now turn to consider the column in the matrix that corresponds to $\omega$. Then $Y_\omega(v) = 1/k$ for each of the $k$ vertices in $P^{-1}(\omega)$ and $Y_\omega(v) = 0$ for all other vertices since if we look for a specific vertex in the graph with property $\omega$ and that is the only information that we know about that sought-after vertex, then it can be any one of the vertices in $P^{-1}(\omega)$ with probability $1/k$.

We are ready to define our notion of privacy.

DEFINITION 2 ($(k, \varepsilon)$-OBFUSCATION). *Let $P$ be a vertex property, $k \geq 1$ be a desired level of obfuscation, and $\varepsilon \geq 0$ be a tolerance parameter. The uncertain graph $\tilde{G}$ is said to $k$-obfuscate a given vertex $v \in G$ with respect to $P$ if the entropy of the distribution $Y_{P(v)}$ over the vertices of $\tilde{G}$ is greater than or equal to $\log_2 k$:*

$$H(Y_{P(v)}) \geq \log_2 k.$$

*The uncertain graph $\tilde{G}$ is a $(k, \varepsilon)$-obfuscation with respect to property $P$ if it $k$-obfuscates at least $(1 - \varepsilon)n$ vertices in $G$ with respect to $P$.*

Namely, given the considered attack scenario, in which the adversary uses a background knowledge of property $P$ of his target vertex, we wish to lower bound the entropy of the distribution it induces over the obfuscated graph vertices by $\log_2 k$ (in similarity to the privacy goal in $k$-anonymity). As for the tolerance parameter $\varepsilon$, it serves the following purpose. Considering the fact that degree sequences in typical social networks have very skewed distribution, trying to obfuscate some very unique vertices (such as Barack Obama or CNN in `twitter` or `Facebook`) is on the one hand hopeless, and on the other hand not necessarily needed: these vertices do not represent "normal" users, and identifying them does not disclose anyone's personal information. In fact, as we will see later, our obfuscation algorithm guarantees that the $\varepsilon$-fraction of vertices for which the privacy requirement is not satisfied can be forced to be taken from some specific sub-population; for example, in the case of degree obfuscation they are vertices with high degree.

EXAMPLE 2. *Consider again the graph in Figure 1. Vertex $v_1$ has degree 3 in the original graph. Thus, in order to check the level of obfuscation of this vertex in the obfuscated graph we have to measure the entropy of the column $\deg = 3$ of Table $Y_\omega(v)$. That entropy is approximately 0.469, which is rather low, meaning that the identity of $v_1$ is not obfuscated enough in the uncertain graph in Figure 1(b). Vertex $v_2$ has degree 1 in the original graph. The entropy of the column $\deg = 1$ is $\approx 1.688 > \log_2 3$. Vertices $v_3$ and $v_4$ have degree 2, and the entropy of the corresponding column is $\approx 1.742 \geq \log_2 3$. Therefore, as three out of four vertices are 3-obfuscated, the graph in Figure 1(b) provides a (3,0.25)-obfuscation for the graph in Figure 1(a).*

## 4. QUANTIFYING THE OBFUSCATION

In this section we describe how to compute the level of obfuscation with regard to the degree property. When $P$ is the degree, $\Omega_P = \{0, \ldots, n-1\}$, and, consequently, the matrix has $n$ rows and $n$ columns. We need to describe how to compute $X_v(\omega)$ for all $v \in \tilde{G}$ and $\omega \in \{0, \ldots, n-1\}$. Once the full matrix $X_v$ is given, it is possible to derive the distributions $Y_\omega$ over the vertices of $\tilde{G}$ for all $\omega \in P(G)$ and then verify the $k$-obfuscation property.

Fix $v \in \tilde{G}$ and let $e_1, \ldots, e_{n-1}$ be the $n-1$ pairs of vertices that include $v$. For each $1 \leq i \leq n-1$, $e_i$ is a Bernoulli random variable that equals 1 with some probability $p_i$. Letting $d_v$ be the random variable corresponding to the degree of $v$, we have

$$d_v = \sum_{i=1}^{n-1} e_i. \qquad (4)$$

Then for each possible degree $\omega \in \Omega_P$ of $v$, we have $X_v(\omega) = \Pr(d_v = \omega)$.

LEMMA 1. *The probability distribution of $d_v$ may be computed exactly in time $O(n^2)$.*

PROOF. Let $d_v^\ell := \sum_{i=1}^{\ell} e_i$ denote the partial sum of the first $\ell$ Bernoulli random variables. We will show that once we have the distribution of $d_v^{\ell-1}$, we can compute that of $d_v^\ell$ in time $O(\ell)$. Hence, the distribution of $d_v = d_v^{n-1}$ can be computed in time $\sum_{\ell=1}^{n-1} O(\ell) = O(n^2)$. Indeed,

$$\Pr(d_v^\ell = j) = \Pr(d_v^{\ell-1} = j-1) \cdot p_\ell + \Pr(d_v^{\ell-1} = j) \cdot (1 - p_\ell).$$

Therefore, computing a single probability in the distribution of $d_v^\ell$ takes constant time (given the full distribution of $d_v^{\ell-1}$), and, consequently, computing the entire distribution of $d_v^\ell$ over all $0 \leq j \leq \ell$ takes time $O(\ell)$. □

It should be noted that since we choose to inject uncertainty only to a subset $E_C$ of pairs of vertices, the sum in Equation (4) is taken only over the pairs of vertices in $E_C$ that include the vertex $v$. Hence, if $d$ is the average degree in $G$, the average number of addends in $d_v$ is $dc$, where $c = |E_C|/|E|$.

In cases where the sum in Equation (4) has a large number of addends, we may adopt an alternative approach. Since $d_v$ is the sum of independent random variables, it may be approximated by the normal distribution $N(\mu, \sigma^2)$, where $\mu = \sum_{i=1}^{n-1} E(e_i) = \sum_{i=1}^{n-1} p_i$ and $\sigma^2 = \sum_{i=1}^{n-1} Var(e_i) = \sum_{i=1}^{n-1} p_i(1-p_i)$ as implied by the Central Limit Theorem [16]. (The Central Limit Theorem becomes effective already for $n \approx 30$; for typical sizes of $n$ in social networks, the normal approximation becomes very accurate.) Specifically, $\Pr(d_v = \omega) \approx \int_{\omega-1/2}^{\omega+1/2} \Phi_{\mu,\sigma}(x)dx$ for $\omega \in \Omega_P = \{0, \ldots, n-1\}$, where

$$\Phi_{\mu,\sigma}(x) = \frac{1}{\sqrt{2\pi\sigma^2}} \cdot e^{-\frac{(x-\mu)^2}{2\sigma^2}}. \qquad (5)$$

## 5. INJECTING UNCERTAINTY

In this section we describe our algorithm, which, given a graph $G$, a desired level of obfuscation $k$, and a tolerance parameter $\varepsilon$, injects a minimal level of uncertainty to the graph so that it becomes $(k, \varepsilon)$-obfuscated with respect to a vertex property $P$.

### 5.1 Overview

As discussed in Section 3, we inject uncertainty in the graph by assigning probabilities to a subset $E_C \subseteq V_2$ of pairs of vertices, such that $|E_C| = c|E|$, for a small constant parameter $c$. The selection of $E_C$ is described in a subsequent section. Once $E_C$ is selected, only the pairs $e \in E_C$ will become uncertain edges in $\tilde{G}$. All other pairs $e \notin E_C$ will be certain non-edges, i.e., $\mathbf{p}(e) = 0$. To establish the uncertainty of each pair $e \in E_C$, we select a random perturbation $r_e \in [0, 1]$. If $e \in E$, it becomes an uncertain edge in $\tilde{G}$ with probability $\mathbf{p}(e) = 1 - r_e$; if $e \in E_C \setminus E$, it becomes an uncertain edge with probability $\mathbf{p}(e) = r_e$.

In order for the uncertain graph $\tilde{G}$ to preserve the characteristics of the original graph $G$, smaller values of the perturbation parameter $r_e$ should be favored. A natural candidate for the generating distribution of $r_e$ is the $[0, 1]$-truncated normal distribution,

$$R_\sigma(r) := \begin{cases} \frac{\Phi_{0,\sigma}(r)}{\int_0^1 \Phi_{0,\sigma}(x)dx} & 0 \leq r \leq 1 \\ 0 & \text{otherwise,} \end{cases} \qquad (6)$$

where $\Phi_{\mu,\sigma}$ is the density function of a Gaussian distribution provided in Equation (5). As the standard deviation $\sigma$ of the normal distribution decreases, a greater mass of $R_\sigma$ will concentrate near $r = 0$ and then the amount of injected uncertainty will be smaller. Thus, small values of $\sigma$ contribute towards better maintaining the characteristics of the original graph, but at the same time they provide lower levels of obfuscation. Larger values of $\sigma$ have the opposite effect.

A key feature of our method is to select judiciously the perturbation $r_e$ for each pair $e = (u, v) \in E_C$, depending on properties of the vertices $u$ and $v$. Hence, the random variable $r_e$ is drawn from $R_{\sigma(e)}$, where the parameter $\sigma(e)$ depends on the vertices that $e$ connects. The perturbation will be larger for edges that connect more unique vertices, which, consequently, require higher levels of uncertainty to "blend in the crowd," and smaller for edges that connect more "typical" vertices.

Additionally, in order to prevent identifying pairs $e \in E_C$ that are true edges in $G$ (by turning every pair $e \in E_C$ to an edge if $\mathbf{p}(e) \geq 0.5$ and to a non-edge otherwise), the perturbation $r_e$ is drawn from the *uniform* distribution in $[0, 1]$, rather than from the distribution $R_\sigma$, for a $q$-fraction of the pairs $e \in E_C$, with $0 < q \ll 1$.

## 5.2 Uniqueness Scores of Vertices

For certain properties of interest, such as degree, the majority of vertices in real-world graphs are already anonymous even without random perturbations. The reason is that for most values of the property $P$ there are many vertices that have that value. Hence, we aim at controlling the amount of applied perturbation, so that larger perturbation is added at vertices that are less anonymized in the original graph. In particular, we suggest to calibrate the perturbation applied to a pair $e = (u, v) \in E_C$ according to the "uniqueness" of the two vertices $u$ and $v$ with respect to the property $P$. Namely, if both $P(u)$ and $P(v)$ are frequent values, then $r_e$ should be very small; on the other hand, if $P(u)$ and $P(v)$ are outlier values, then $r_e$ should be higher. We proceed to explain our method in detail.

Let $P : V \to \Omega_P$ be a property defined on the set of vertices $V$. Further, consider a distance function $d$ between values in the range $\Omega_P$ of $P$. So, for each pair of values, $\omega, \omega' \in \Omega_P$, a distance $d(\omega, \omega') \geq 0$ is defined. For example, for the degree property $P_1$, the distance $d$ is the modulus of the difference of two degrees, while for the radius-one subgraph property (P3), the distance $d$ is the edit distance between two subgraphs.

DEFINITION 3. *Let $P : V \to \Omega_P$ be a property on the set of vertices $V$ of the graph $G$, let $d$ be a distance function on $\Omega_P$, and let $\theta > 0$ be a parameter. Then the $\theta$-commonness of the property value $\omega \in \Omega_P$ is $C_\theta(\omega) := \sum_{v \in V} \Phi_{0,\theta}(d(w, P(v)))$, while the $\theta$-uniqueness of $\omega \in \Omega_P$ is $U_\theta(\omega) := \frac{1}{C_\theta(\omega)}$.*

In the above definition the function $\Phi$ is the Gaussian distribution given by Equation (5). The commonness of the property value $\omega$ is a measure of how typical is the value $\omega$ among the vertices of the graph. It is obtained as a weighted average over all other property values $\omega'$, where the weight decays exponentially as a function of the distance between $\omega$ and $\omega'$. The uniqueness is the inverse of the commonness. It should be noted that the commonness and uniqueness are meaningful only as relative measures, as they allow to assess how one property value is more common, or more unique, in $G$ than another property value.

Commonness and uniqueness scores depend on the parameter $\theta$, which determines the decay rate of the average weights as a function of the distance. We set $\theta = \sigma$ as larger amounts of uncertainty imply that property values may be spread on larger domains of $\Omega_P$ due to injecting uncertainty.

## 5.3 The Obfuscation Algorithm

Our algorithm for computing a $(k, \varepsilon)$-obfuscation of a graph with respect to a vertex property $P$ is outlined as Algorithm 1. Targeting for high utility, the algorithm aims at injecting the minimal amount of uncertainty needed to achieve the required obfuscation. Computing the minimal

---

**Algorithm 1** $(k, \varepsilon)$-obfuscation

**Input**: $G = (V, E)$, vertex property $P$, obfuscation level $k$, tolerance $\varepsilon$, size multiplier $c$, and white noise level $q$.
**Output**: A $(k, \varepsilon)$-obfuscation $\tilde{G}$ of $G$ with respect to $P$.
1: $\sigma_\ell \leftarrow 0$
2: $\sigma_u \leftarrow 1$
3: **repeat**
4: $\quad \langle \tilde{\varepsilon}, \tilde{G} \rangle \leftarrow \mathsf{GenerateObfuscation}(G, \sigma_u, P, k, \varepsilon, c, q)$
5: $\quad$ **if** $\tilde{\varepsilon} = \infty$ **then** $\sigma_u \leftarrow 2\sigma_u$
6: **until** $\tilde{\varepsilon} \neq \infty$
7: $\tilde{G}_{found} \leftarrow \tilde{G}$
8: **while** $\sigma_\ell + \delta < \sigma_u$ **do**
9: $\quad \sigma \leftarrow (\sigma_\ell + \sigma_u)/2$
10: $\quad \langle \tilde{\varepsilon}, \tilde{G} \rangle \leftarrow \mathsf{GenerateObfuscation}(G, \sigma_u, P, k, \varepsilon, c, q)$
11: $\quad$ **if** $\tilde{\varepsilon} = \infty$ **then** $\sigma_\ell \leftarrow \sigma$
12: $\quad$ **else** $\tilde{G}_{found} \leftarrow \tilde{G}$; $\sigma_u \leftarrow \sigma$
13: **return** $\tilde{G}_{found}$

---

amount of uncertainty is achieved via a binary search on the value of the uncertainty parameter $\sigma$.

The binary-search flow of Algorithm 1 is determined by the function GenerateObfuscation, which is shown as Algorithm 2. The function GenerateObfuscation returns a pair $\langle \tilde{\varepsilon}, \tilde{G} \rangle$ where $\tilde{\varepsilon} = \infty$ or $0 \leq \tilde{\varepsilon} \leq \varepsilon$. In the first case, the function could not find a $(k, \varepsilon)$-obfuscation with the given uncertainty parameter. In the latter case, $\tilde{G}$ is a $(k, \tilde{\varepsilon})$-obfuscation of $G$ with respect to $P$, and thus, also a $(k, \varepsilon)$-obfuscation.

The obfuscation algorithm starts with an initial guess of an upper bound $\sigma_u$, which is iteratively doubled until a $(k, \varepsilon)$-obfuscated graph is found. Then, the binary-search process is performed using $\sigma_\ell = 0$ as the lower bound, and the upper bound $\sigma_u$ that was found. The binary search terminates when the search interval is sufficiently short, and the algorithm outputs the best $(k, \varepsilon)$-obfuscation found (i.e., the last one that was successfully generated, because it will be the one obtained with the smallest $\sigma$).

The function GenerateObfuscation (Algorithm 2) aims at finding a $(k, \varepsilon)$-obfuscation of $G$ using a given standard deviation parameter $\sigma$. First, it computes the $\sigma$-uniqueness level $U_\sigma(P(v))$ for each vertex $v \in G$. The more unique a vertex is, the harder it is to obfuscate it. Hence, in order to use the "uncertainty budget" $\sigma$ in the most efficient way, the algorithm performs the following two pre-processing steps.

(Line 2): Since it is allowed not to obfuscate $\varepsilon |V|$ of the vertices, the algorithm selects the set $H$ of $\lceil \frac{\varepsilon}{2} |V| \rceil$ vertices with largest uniqueness scores, which are the vertices that would require the largest amount of uncertainty, and excludes them from the subsequent obfuscation efforts. In later steps, the algorithm will inject uncertainty only to edges that are not adjacent to any of the vertices in $H$. (The algorithm could also receive $H$, or part of $H$, as an input, instead of fully selecting it on its own.)

(Line 3): The set of vertices not in $H$ will need to be obfuscated. To obfuscate more unique vertices, higher uncertainty is necessary. Thus, edges need to be sampled with higher probability if they are adjacent to unique vertices. In order to handle this sampling process, our algorithm assigns a probability $Q(v)$ to every $v \in V$, which is proportional to the uniqueness level $U_\sigma(P(v))$ of $v$.

After that, the search for a $(k, \varepsilon)$-obfuscation starts: since the algorithm is randomized and there is a non-zero prob-

1380

## Algorithm 2 GenerateObfuscation

**Input**: $G = (V, E), P, k, \varepsilon, c, q$, and standard deviation $\sigma$.
**Output**: A pair $\langle \tilde{\varepsilon}, \tilde{G} \rangle$, where $\tilde{G}$ is a $(k, \tilde{\varepsilon})$-obfuscation (with $\tilde{\varepsilon} < \varepsilon$), or $\tilde{\varepsilon} = \infty$ if a $(k, \varepsilon)$-obfuscation was not found.

1: **for all** $v \in V$ **compute** the $\sigma$-uniqueness $U_\sigma(P(v))$
2: $H \leftarrow$ the set of $\lceil \frac{\varepsilon}{2} |V| \rceil$ vertices with largest $U_\sigma(P(v))$
3: **for all** $v \in V$ **do** $Q(v) \leftarrow U_\sigma(P(v))/\sum_{u \in V} U_\sigma(P(u))$
4: $\tilde{\varepsilon} \leftarrow \infty$
5: **for** $t$ times **do**
6: $\quad E_C \leftarrow E$
7: $\quad$ **repeat**
8: $\quad\quad$ randomly pick a vertex $u \in V \setminus H$ according to $Q$
9: $\quad\quad$ randomly pick a vertex $v \in V \setminus H$ according to $Q$
10: $\quad\quad$ **if** $(u, v) \in E$ **then** $E_C \leftarrow E_C \setminus \{(u, v)\}$
11: $\quad\quad$ **else** $E_C \leftarrow E_C \cup \{(u, v)\}$
12: $\quad$ **until** $|E_C| = c|E|$
13: $\quad$ **for all** $e \in E_C$ **do**
14: $\quad\quad$ compute $\sigma(e)$
15: $\quad\quad$ draw $w$ uniformly at random from $[0, 1]$
16: $\quad\quad$ **if** $w < q$
17: $\quad\quad$ **then** draw $r_e$ uniformly at random from $[0, 1]$
18: $\quad\quad$ **else** draw $r_e$ from the random distribution $R_{\sigma(e)}$
19: $\quad\quad$ **if** $e \in E$ **then** $\mathbf{p}(e) \leftarrow 1 - r_e$ **else** $\mathbf{p}(e) \leftarrow r_e$
20: $\quad \varepsilon' \leftarrow |\{v \in V : \text{not } k\text{-obfuscated by } G' = (V, \mathbf{p})\}|/|V|$
21: $\quad$ **if** $\varepsilon' \leq \varepsilon$ **and** $\varepsilon' < \tilde{\varepsilon}$ **then** $\tilde{\varepsilon} \leftarrow \varepsilon'$; $\tilde{G} \leftarrow G'$
22: **return** $\langle \tilde{\varepsilon}, \tilde{G} \rangle$

ability of failure, $t$ attempts to find a $(k, \varepsilon)$-obfuscation are performed (Lines 5-22; in our experiments we used $t = 5$).

Each attempt begins by randomly selecting a subset $E_C \subseteq V_2$, which will be subjected to uncertainty injection (Lines 6-12). The set $E_C$, whose target size is $|E_C| = c|E|$, is initialized to be $E$ (Line 6). Then, the algorithm randomly selects two distinct vertices $u$ and $v$, according to the probability distribution $Q$, such that none of them is in $H$ (Lines 8-9). The pair of vertices $(u, v)$ is removed from $E_C$ if it is an edge, or added to $E_C$ otherwise (Lines 10-11). The process is repeated until $E_C$ reaches the required size $c|E|$. Since in typical graphs, the number of non-edges is significantly larger than the number of edges, i.e., $|E| \ll |V_2|/2$, the loop in Lines 7-12 ends very quickly, for small values of $c$, and the resulting set $E_C$ includes most of the edges in $E$.

Next, in Line 14, we redistribute the uncertainty levels among all pairs $e \in E_C$ in proportion to their uniqueness levels. Specifically, we define for each $e = (u, v) \in E_C$ its $\sigma$-uniqueness level,

$$U_\sigma(e) := \frac{U_\sigma(P(u)) + U_\sigma(P(v))}{2},$$

and then set

$$\sigma(e) = \sigma |E_C| \cdot \frac{U_\sigma(e)}{\sum_{e' \in E_C} U_\sigma(e')}, \quad (7)$$

so that the average of $\sigma(e)$ over all $e \in E_C$ equals $\sigma$.

Given the edge uncertainty levels, $\sigma(e)$, we select for each pair of vertices $e \in E_C$ a random perturbation $r_e$. For the majority of the pairs (an $(1 - q)$-fraction, where the input parameter $q$ is small) we select $r_e$ from the random distribution $R_{\sigma(e)}$ (see Equation (6)). For the remaining $q$-fraction of pairs we select $r_e$ from the uniform distribution on $[0, 1]$. If $e$ is an actual edge ($e \in E$), it turns into an uncertain edge in $\tilde{G}$ with associated probability of $\mathbf{p}(e) = 1 - r_e$. If $e$ is a non-edge in $G$ ($e \in E_C \setminus E$), it turns into an uncertain edge in $\tilde{G}$ with probability $\mathbf{p}(e) = r_e$ (Line 19).

If the algorithm finds a $(k, \varepsilon)$-obfuscated graph in one of its $t$ trials, it returns the obfuscated graph with minimal $\varepsilon$. If, on the other hand, all $t$ attempts fail, the algorithm indicates the failure by returning $\tilde{\varepsilon} = \infty$.

## 6. UTILITY OF THE UNCERTAIN GRAPH

In order to prove the practical relevance of our proposal, we need to show that: (1) the uncertain graph maintains high utility, i.e., it is highly similar to the original graph in terms of characteristic properties; and (2) the computation of these properties can be carried out in reasonable time.

In the rest of this section, we discuss several graph statistics and show how to compute them in uncertain graphs. In our experimental assessment, we use those statistics to evaluate the utility of the proposed graph obfuscation.

Further evidence to the usefulness of publishing an uncertain graph is provided by the many recent papers on mining and querying uncertain graphs [14, 15, 24, 36, 37, 38].

### 6.1 Sampling

Given a standard (certain) graph $G$, let $S[G]$ be the value of a statistical measure $S$ for $G$. Examples of such a statistical measure $S$ are the average degree, the diameter, the clustering coefficient of $G$, and so on. In order to define the value of $S$ in an uncertain graph $\tilde{G} = (V, \mathbf{p})$, the most natural choice is to consider the *expected value* of $S[\tilde{G}]$, namely,

$$E(S[\tilde{G}]) = \sum_{W \in \mathcal{W}(\tilde{G})} \Pr(W) \cdot S(W), \quad (8)$$

where $\Pr(W)$ is given in Equation (1). While for some statistics it is possible to compute the expected value in Equation (8) without explicitly performing a summation over the exponential number of possible worlds (as we will see in Section 6.2), for other statistics such a computation remains infeasible. Hence, we have to resort to approximation by sampling. Namely, we sample a subset of possible worlds $\mathcal{W}' \subseteq \mathcal{W}(\tilde{G})$ according to the distribution induced by the probabilities $\Pr(W)$, and then take the average $\overline{S}$ of the statistic $S$ in the sampled worlds as an approximation of $E(S[\tilde{G}])$:

$$\overline{S} := \frac{1}{|\mathcal{W}'|} \sum_{W \in \mathcal{W}'} S(W). \quad (9)$$

Sampling a possible world according to the distribution $\Pr(W)$ is carried out by sampling independently each edge $e$ with probability $\mathbf{p}(e)$.

The following lemma provides a probabilistic error bound for approximating the expected value by an average over a number of sampled worlds.

LEMMA 2. *Let $\tilde{G} = (V, \mathbf{p})$ be an uncertain graph and assume that $S$ is a graph statistic that satisfies $a \leq S \leq b$. Let $r = |\mathcal{W}'|$ denote the number of sampled worlds and $\overline{S}$ be the average of the statistic $S$ over those worlds, Equation (9). Then for every $\varepsilon > 0$,*

$$\Pr(|E(S[\tilde{G}]) - \overline{S}| \geq \varepsilon) \leq 2 \exp\left(-\frac{2\varepsilon^2 r}{(b-a)^2}\right). \quad (10)$$

PROOF. Let $\mathcal{W}' = \{W_i\}_{1 \leq i \leq r}$ be the set of $r$ graphs that were sampled from $\tilde{G} = (V, \mathbf{p})$. Then $S_i = S[W_i]$, $1 \leq i \leq r$, are independent and identically distributed random variables. Since $E(S_i) = E(S[\tilde{G}])$ for all $1 \leq i \leq r$, it follows that also $E(\overline{S}) = E(S[\tilde{G}])$. Hence, inequality (10) follows directly from Hoeffding's inequality [13]. □

COROLLARY 1. *For given error bound $\varepsilon$ and probability of failure $\delta$, we have $\Pr(|E(S[\tilde{G}]) - \overline{S}| \geq \varepsilon) \leq \delta$, provided that $r \geq \frac{1}{2}\left(\frac{b-a}{\varepsilon}\right)^2 \ln\left(\frac{2}{\delta}\right)$.*

In the next section, we define a number of scalar and vector statistics of interest; when possible, we also provide an explicit computation of $E(S[\tilde{G}])$.

## 6.2 Statistics Based on Degree

Let $d_1, \ldots, d_n$ denote the degree sequence in a graph $G$. The statistic $S$ is called a degree-based statistic if $S = F(d_1, \ldots, d_n)$ for some function $F$. Examples of such statistics are:

- *Number of edges*: $S_{\mathrm{NE}} = \frac{1}{2} \sum_{v \in V} d_v$.
- *Average degree*: $S_{\mathrm{AD}} = \frac{1}{n} \sum_{v \in V} d_v$.
- *Maximal degree*: $S_{\mathrm{MD}} = \max_{v \in V} d_v$.
- *Degree variance*:[1] $S_{\mathrm{DV}} = \frac{1}{n} \sum_{v \in V} (d_v - S_{\mathrm{AD}})^2$.

When $\tilde{G}$ is an uncertain graph, $d_1, \ldots, d_n$ are random variables. If $F$ is a linear function, then we have

$$E(S[\tilde{G}]) = E(F(d_1, \ldots, d_n)) = F(E(d_1), \ldots, E(d_n)). \quad (11)$$

Hence, since the expected degree of a vertex $v \in V$ is equal to the sum of probabilities of its adjacent edges, the computation of the expected statistic is easy, in the case of a linear function. As the first two examples above, $S_{\mathrm{NE}}$ and $S_{\mathrm{Ad}}$, correspond to a linear function $F$, we have:

$$E(S_{\mathrm{NE}}[\tilde{G}]) = E\left(\frac{1}{2} \sum_{v \in V} d_v\right) = \frac{1}{2} \sum_{v \in V} \sum_{u \in V \setminus v} \mathbf{p}(u, v) = \sum_{e \in V_2} \mathbf{p}(e),$$

and

$$E(S_{\mathrm{AD}}[\tilde{G}]) = E\left(\frac{1}{n} \sum_{v \in V} d_v\right) = \frac{1}{n} \sum_{v \in V} \sum_{u \in V \setminus v} \mathbf{p}(u, v) = \frac{2}{n} \sum_{e \in V_2} \mathbf{p}(e).$$

Things are less simple when $F$ is non-linear, since then Equation (11) does not hold. This is the case with the latter two examples — the maximal degree ($F = \max$) and the degree variance ($F$ is quadratic). For these statistics we adopt the sampling approach described in the previous section. Since the maximal degree is at most $n-1$, the statistic $S_{\mathrm{MD}}$ satisfies Corollary 1 with $a = 0$ and $b = n - 1$. Similarly, the statistic $S_{\mathrm{DV}}$ satisfies Corollary 1 with $a = 0$ and $b = (n-1)^2$. It should also be noted that we can compute $E(S_{\mathrm{DV}}[\tilde{G}])$ precisely. However, the cost of evaluating the corresponding formulas, which we omit herein, is quadratic in the number of vertices.

We proceed to describe two additional statistics that are based on the degree distribution. In the following we use $\Delta(d)$, with $0 \leq d \leq n - 1$, to denote the fraction of vertices in the graph $G$ that have degree $d$.

---
[1] This is one of the measures of graph heterogeneity [27].

The first statistic, denoted by $S_{\mathrm{PL}}$, is the power-law exponent of the degree distribution. For this statistic, we assume that the degree distribution follows a power law, $\Delta(d) \sim d^{-\gamma}$, and $S_{\mathrm{PL}}$ is an estimate of $-\gamma$. In our experiments, we focused on higher degrees where the power law fits better, and we fitted the exponent ignoring smaller degrees.

The second statistic is the degree distribution itself, $S_{\mathrm{DD}} := (\Delta(0), \Delta(1), \ldots, \Delta(n-1))$. As opposed to all previous statistics, which were scalar, this one is a vector. In fact, each of the previous statistics may be derived from the degree distribution. To approximate $S_{\mathrm{DD}}[G]$ we adopt once more the sampling approach: for every degree $d$, we approximate $\Delta(d)$ by the average $\overline{\Delta}(d)$ obtained over the sampled possible worlds.

## 6.3 Statistics Based on Shortest-path Distance

Other interesting measures characterizing a graph are those based on the shortest-path distance between pairs of vertices. Computing distance distributions on large graphs is far from trivial, as explained in the survey of Kang et al. [17]. While exact solutions using breadth-first search or Floyd's algorithm are out of question, there is still no consensus in the research community on which approximate technique is best [9]. Some methods are based on sampling, for example, performing a breadth-first search from a selected set of vertices [6, 18], and other are based on information diffusion [3, 17, 23]. While the former are simpler to implement, diffusion-based techniques have the advantage of being more general (they are natively designed for directed graphs, while most sampling methods only work for undirected ones) and scale more gracefully.

Defining the distance between pairs of vertices in uncertain graphs is not an easy task since, typically, the corresponding ensemble of possible worlds will include disconnected instances; in such disconnected possible worlds, some of the pairwise distances are infinite [24]. We directly avoid this problem by defining the distance-based measures $S$ only on pairs of vertices that are path-connected.

We consider five measures:

- *Average distance*: $S_{\mathrm{APD}}$ is the average distance among all pairs of vertices that are path-connected.

- *Effective diameter*: $S_{\mathrm{EDiam}}$ is the 90-*th* percentile distance among all path-connected pairs of vertices, i.e., the minimal value for which 90% of the finite pairwise distances in the graph are no larger than. In our experiments, we used the variant that linearly interpolates between the 90-*th* percentile and the successive integer.

- *Connectivity length*: The statistic $S_{\mathrm{CL}}$ is defined as the harmonic mean of all pairwise distances in the graph, $S_{\mathrm{CL}} = \frac{n(n-1)}{2} \left(\sum_{(u,v) \in V_2} \frac{1}{\mathrm{dist}(u,v)}\right)^{-1}$ [20]. Note that by taking $\frac{1}{\mathrm{dist}(u,v)} = 0$ for non path-connected pairs $(u, v)$, the connectivity length can be defined as the average over all vertex pairs, independently on whether they lie in the same connected component.

- *Distribution of pairwise distances*: $S_{\mathrm{PDD}}$ is the distribution of pairwise distances in the graph, where $S_{\mathrm{PDD}}[k]$ is the number of pairs of vertices whose distance equals $k$, for $1 \leq k \leq n - 1$, and $S_{\mathrm{PDD}}[\infty]$ is the number of pairs of vertices that are not path-connected.

- *Diameter*: $S_{\mathrm{Diam}}$ is the maximum distance among all path-connected pairs of vertices.

For computing the above measures we rely on sampling. It is easy to see that Lemma 2 and Corollary 1 hold for each of those statistics with $a = 1$ and $b = n - 1$.

To estimate the distance distribution in a given (certain) graph, we use HyperANF [3], a diffusion-based algorithm that provides a good tradeoff between accuracy guarantees and execution time. As the algorithm is probabilistic, the results that it gives may drift from the real ones, depending on the number of registers used for the evaluation. Such drifts affect the variance over the value obtained for each point of the distance distribution. To limit the effect of such probabilistic drifts, we repeat the execution of HyperANF and used jackknifing [26] to infer the standard error of the statistics that we compute; in our experiments this error ranges between 0.2% and 2%.

While the HyperANF approach is viable for the first four statistics described above, it falls short in estimating the diameter. Exact diameter estimation is difficult and even heuristic methods such as [9] would be too inefficient to be executed on many sampled worlds. As a result, we focus on estimating a lower bound $S_{\text{DiamLB}}$ for $S_{\text{Diam}}$: such a lower bound is computed as the largest distance $t$ for which the approximate distance distribution computed by HyperANF is nonzero; i.e., it is the largest distance $t$ for which we estimate that there is at least one pair of vertices of distance $t$ from each other.

### 6.4 Clustering Coefficient

The clustering coefficient $S_{\text{CC}}$ measures the extent to which the edges of the graph "close triangles." More formally, given a graph $G$, let $T_3[G]$ be the number of cliques of size 3 in the graph $G$, and $T_2[G]$ be the number of connected triplets. The clustering coefficient $S_{\text{CC}}[G]$ of a graph $G$ is then defined as $S_{\text{CC}}[G] = T_3[G]/T_2[G]$. Since $T_3[G] \leq T_2[G]$, the clustering coefficient is a number between 0 and 1.

EXAMPLE 3. *Let $K_3$ be the complete graph on three vertices. Then $T_3[K_3] = 1$ and $T_2[K_3] = 1$. Hence, $S_{\text{CC}}[K_3] = 1$. Consider next the graph $G$ on three vertices $u, v, w$ with two edges only — $(u, v)$ and $(u, w)$. Then $T_3[G] = 0$ and $T_2[G] = 1$, whence $S_{\text{CC}}[G] = 0$.*

Given an uncertain graph $\tilde{G}$, we can estimate the expected clustering coefficient $E(S_{\text{CC}}[\tilde{G}])$ by sampling (see Section 6.1). Since the clustering coefficient takes values between 0 and 1, we can apply Lemma 2 with $a = 0$ and $b = 1$. Thus, we can estimate $E(S_{\text{CC}}[\tilde{G}])$ within an error of at most $\varepsilon$ and probability of success at least $1 - \delta$ by sampling at least $r = \frac{1}{2\varepsilon^2} \ln(\frac{2}{\delta})$ possible worlds.

### 7. EXPERIMENTAL ASSESSMENT

The objective of our experimental assessment is to show that the proposed technique is able to provide the required obfuscation levels while maintaining high data utility. In particular, we set the following concrete subgoals. For given values of $k$ and $\varepsilon$, we want to assess:

1. the level of noise (specified by the value of $\sigma$) needed to achieve $(k, \varepsilon)$-obfuscation;
2. the running time of the obfuscation algorithm;
3. the error in the statistics of the obfuscated graph with respect to the original graph;
4. how the proposed method compares with random-perturbation methods for the same levels of obfuscation.

**Table 2: Values of $\sigma$ that yielded a $(k, \varepsilon)$-obfuscation obtained by Alg. 1. In all cases $q = 0.01$ and $c = 2$, except for the two cases marked (*) where $c = 3$.**

| Dataset | $k$ | $\varepsilon = 10^{-3}$ | $\varepsilon = 10^{-4}$ |
|---|---|---|---|
| dblp | 20 | $5.9605 \cdot 10^{-8}$ | $1.6153 \cdot 10^{-5}$ |
| | 60 | $2.9802 \cdot 10^{-7}$ | $3.2206 \cdot 10^{-3}$ |
| | 100 | $1.8775 \cdot 10^{-5}$ | $1.0711 \cdot 10^{-2}$ |
| flickr | 20 | $2.2948 \cdot 10^{-5}$ | $2.6343 \cdot 10^{-2}$ |
| | 60 | $1.0397 \cdot 10^{-3}$ | $7.3275 \cdot 10^{-2}$ (*) |
| | 100 | $5.8624 \cdot 10^{-3}$ | $2.9273 \cdot 10^{-1}$ (*) |
| Y360 | 20 | $5.9605 \cdot 10^{-8}$ | $5.9605 \cdot 10^{-8}$ |
| | 60 | $5.9605 \cdot 10^{-8}$ | $1.0133 \cdot 10^{-6}$ |
| | 100 | $5.9605 \cdot 10^{-8}$ | $1.1146 \cdot 10^{-5}$ |

**Table 3: Computation (real) time in edges/sec.**

| Dataset | $k$ | $\varepsilon = 10^{-3}$ | $\varepsilon = 10^{-4}$ |
|---|---|---|---|
| dblp | 20 | 1069.34 | 1550.78 |
| | 60 | 1000.64 | 1279.39 |
| | 100 | 888.908 | 1166.87 |
| flickr | 20 | 1004.93 | 926.45 |
| | 60 | 1019.05 | 300.39 (*) |
| | 100 | 862.155 | 271.84 (*) |
| Y360 | 20 | 2113.51 | 1900.32 |
| | 60 | 1762.21 | 1665.80 |
| | 100 | 1643.84 | 1664.75 |

For our experiments, we use three large real-world datasets. dblp is a co-authorship graph extracted from a recent snapshot of the DBLP database considering only journal publications.[2] Vertices represent authors, and there is an undirected edge between two authors if they have authored a journal paper together.

flickr is a popular online community for sharing photos, with millions of users.[3] In addition to many photo-sharing facilities, users are creating a social network by explicitly marking other users as their *contacts*.

Y360: Yahoo! 360 was a social-networking and personal-communication portal. In the Y360 dataset, edges represents the friendship relationship among users.

The graphs sizes vary from 226 413 vertices of dblp, 588 166 of flickr, to 1 226 311 of Y360, with different densities; Y360 is the largest but also the sparsest dataset. The main statistics (as defined in Section 6) of the three datasets are reported in Table 4.

### 7.1 Parameter Tuning and Running Time

In our first set of experiments, we considered three obfuscation levels, $k \in \{20, 60, 100\}$, and two possible tolerance values, $\varepsilon \in \{10^{-3}, 10^{-4}\}$. We experimented with different values for $q$ and $c$ (with $q \in \{0.01, 0.05, 0.1\}$ and $c \in \{2, 3\}$), but here we present only the case $q = 0.01$ and $c = 2$ (except for two instances that will be discussed below). In Table 2, we report the minimal values of $\sigma$, as found by Algorithm 1, that yielded a $(k, \varepsilon)$-obfuscation for given values of $k$ and $\varepsilon$.

As expected, larger $k$ or smaller $\varepsilon$ required larger values of $\sigma$, because more noise was needed in order to reach the desired level of obfuscation. In some cases, Algorithm 1 failed to find a proper upper bound for $\sigma$ in the loop in

---

[2] http://www.informatik.uni-trier.de/~ley/db/
[3] http://www.flickr.com/

Table 4: The sample mean on a sample of size 100, with $\varepsilon = 10^{-4}$. The last column is the average (over all statistics) of the relative absolute difference between the sample mean and the real value of the statistics.

| | graph | $S_{\text{NE}}$ | $S_{\text{AD}}$ | $S_{\text{MD}}$ | $S_{\text{DV}}$ | $S_{\text{PL}}$ | $S_{\text{APD}}$ | $S_{\text{DiamLB}}$ | $S_{\text{EDiam}}$ | $S_{\text{CL}}$ | $S_{\text{CC}}$ | rel.err. |
|---|---|---|---|---|---|---|---|---|---|---|---|---|
| dblp | real | 716 460 | 6.33 | 238 | 76.79 | −0.046 | 7.34 | 25 | 8.78 | 6.96 | 0.38 | |
| | $k = 20$ | 713 952 | 6.31 | 233 | 76.18 | −0.046 | 7.01 | 22.59 | 7.16 | 6.68 | 0.35 | 0.049 |
| | $k = 60$ | 735 766 | 6.50 | 652 | 122.8 | −0.014 | 6.05 | 20.52 | 6.29 | 5.76 | 0.23 | 0.429 |
| | $k = 100$ | 754 776 | 6.67 | 975 | 187.6 | −0.008 | 5.67 | 19.12 | 6.00 | 5.41 | 0.16 | 0.705 |
| flickr | real | 5 801 442 | 19.73 | 6 660 | 6 200 | −0.002 | 5.03 | 21 | 5.43 | 4.80 | 0.12 | |
| | $k = 20$ | 5 921 470 | 20.14 | 5 847 | 6 924 | −0.002 | 4.84 | 20.51 | 4.80 | 4.64 | 0.05 | 0.112 |
| | $k = 60$ | 6 944 481 | 23.61 | 4 534 | 12 847 | −0.002 | 4.59 | 17.66 | 4.47 | 4.42 | 0.04 | 0.322 |
| | $k = 100$ | 7 640 446 | 25.98 | 6 121 | 18 438 | −0.001 | 4.50 | 16.81 | 4.33 | 4.37 | 0.06 | 0.415 |
| Y360 | real | 2 618 645 | 4.27 | 258 | 112.6 | −0.027 | 8.21 | 31 | 8.94 | 7.77 | 0.04 | |
| | $k = 20$ | 2 605 027 | 4.25 | 257 | 109.5 | −0.028 | 8.06 | 31.53 | 9.19 | 7.66 | 0.03 | 0.026 |
| | $k = 60$ | 2 605 952 | 4.25 | 256 | 110.0 | −0.028 | 8.05 | 30.04 | 8.95 | 7.64 | 0.03 | 0.025 |
| | $k = 100$ | 2 609 937 | 4.26 | 259 | 111.9 | −0.027 | 8.01 | 31.64 | 8.99 | 7.60 | 0.03 | 0.023 |

Table 5: The relative sample standard error of the mean (SEM) on a sample of size 100, with $\varepsilon = 10^{-4}$ (the other parameters are set as in Table 2). For every statistics, the value shown is the sample standard deviation, divided by the square root of the sample size and normalized by the sample mean. The last column is the average of the relative sample standard errors over all of the statistics.

| | $k$ | $S_{\text{NE}}$ | $S_{\text{AD}}$ | $S_{\text{MD}}$ | $S_{\text{DV}}$ | $S_{\text{PL}}$ | $S_{\text{APD}}$ | $S_{\text{DiamLB}}$ | $S_{\text{EDiam}}$ | $S_{\text{CL}}$ | $S_{\text{CC}}$ | average |
|---|---|---|---|---|---|---|---|---|---|---|---|---|
| dblp | 20 | 0.00010 | 0.00010 | 0.0120 | 0.00100 | 0.0110 | 0.0040 | 0.041 | 0.10 | 0.020 | 0.013 | 0.019 |
| | 60 | 0.00024 | 0.00024 | 0.0260 | 0.00350 | 0.0170 | 0.0035 | 0.058 | 0.16 | 0.019 | 0.018 | 0.028 |
| | 100 | 0.00029 | 0.00029 | 0.0170 | 0.00430 | 0.0170 | 0.0033 | 0.055 | 0.15 | 0.018 | 0.024 | 0.027 |
| flickr | 20 | 0.00016 | 0.00016 | 0.0067 | 0.00074 | 0.0037 | 0.0036 | 0.060 | 0.15 | 0.016 | 0.045 | 0.028 |
| | 60 | 0.00018 | 0.00018 | 0.0100 | 0.00068 | 0.0030 | 0.0039 | 0.084 | 0.17 | 0.018 | 0.054 | 0.033 |
| | 100 | 0.00017 | 0.00017 | 0.0064 | 0.00059 | 0.0032 | 0.0039 | 0.082 | 0.18 | 0.018 | 0.035 | 0.032 |
| Y360 | 20 | 0.00004 | 0.00004 | 0.0024 | 0.00025 | 0.0035 | 0.0036 | 0.043 | 0.13 | 0.021 | 0.045 | 0.027 |
| | 60 | 0.00004 | 0.00004 | 0.0049 | 0.00031 | 0.0032 | 0.0046 | 0.051 | 0.15 | 0.021 | 0.061 | 0.031 |
| | 100 | 0.00005 | 0.00005 | 0.0120 | 0.00044 | 0.0044 | 0.0035 | 0.052 | 0.16 | 0.018 | 0.057 | 0.032 |

Lines 3-6. In those cases, increasing the parameter $c$ to 3 resolved the problem.

The obfuscation algorithm was implemented in Java and run on an Intel Xeon X5660 CPUs, 2.80 GHz, 12 MB cache size. Table 3 reports the running times (expressed in edges per second) of the same experiments for which we reported in Table 2 the values of $\sigma$. As explained above, we used in all cases $q = 0.01$ and $c = 2$, except for the two cases marked by (∗) in which $c = 3$. We note that using smaller values of $c$ has the benefit of keeping the graph size under control; such a benefit is of special importance for large networks. Smaller values of $c$ also reduce the runtime of Algorithm 2, where the main loop (Lines 13-19) is over $c|E|$ edges. This effect is evident in Table 3, where the performance drops substantially in the two cases where $c = 3$. As expected, the performance slightly decreases when $k$ increases or $\varepsilon$ decreases, due to the increased efforts to achieve a higher obfuscation level. We note that the smaller computation times required for Y360 are due to the fact that this dataset turns out to be easier to obfuscate than the others (as witnessed also by the small final values of $\sigma$ as reported in Table 2).

The parameter $q$ just introduces some amount of "white noise" in the graph. Using higher values of $q$ enhances obfuscation but it also reduces the utility of the final released graph. Due to space limitations, we present only results for $q = 0.01$. A more elaborated set of plots, for different settings of $q$ and other obfuscation parameters, will be given in an extended version of this paper[4].

### 7.2 Data Utility

Next, we computed statistics of interest on the obfuscated graphs, using the sampling method (Section 6.1).[5] For every obfuscated graph, we sampled 100 possible worlds and for each of them we computed all the scalar statistics listed above. The mean values obtained are shown in Table 4. Those values are very concentrated, as witnessed by Table 5, that reports the relative sample standard error of the mean (also called SEM; it is obtained as the sample standard deviation divided by the square root of the sample size and normalized by the sample mean); the last column reports the average computed over all the statistics. As can be seen, all statistics are very well concentrated; on average, the fluctuations for all statistics are of about 3% (last column of Table 5), but most of them (see, for example, $S_{\text{NE}}$ or $S_{\text{AD}}$) exhibit a much higher level of concentration. There is a weak dependence on $k$ and also on $\varepsilon$ (the latter dependence is not shown here).

We proceed to comparing the sample mean of the statistics obtained with their real values on the original graph (see again Table 4). The quality of the estimation decreases when obfuscation becomes larger: in the last column of the table, we computed the average statistical error over all scalar statistics, that is, the relative absolute difference between the estimate and the real value. With small values of $k$, e.g., $k = 20$, the error is always well below 15%; larger values of $k$ introduce larger errors, up to 70.5% when $k = 100$ in the dblp dataset. Observe that some statistics (e.g., degree variance or clustering coefficient) are more affected by error than others.

---

[4]A complete set of plots, along with the code of Algorithm 1, is available at http://boldi.dsi.unimi.it/obfuscation/.

[5]For $S_{\text{NE}}$ and $S_{\text{AD}}$ we use the exact formulas (Sec. 6.2). The results are almost identical to those obtained by sampling.

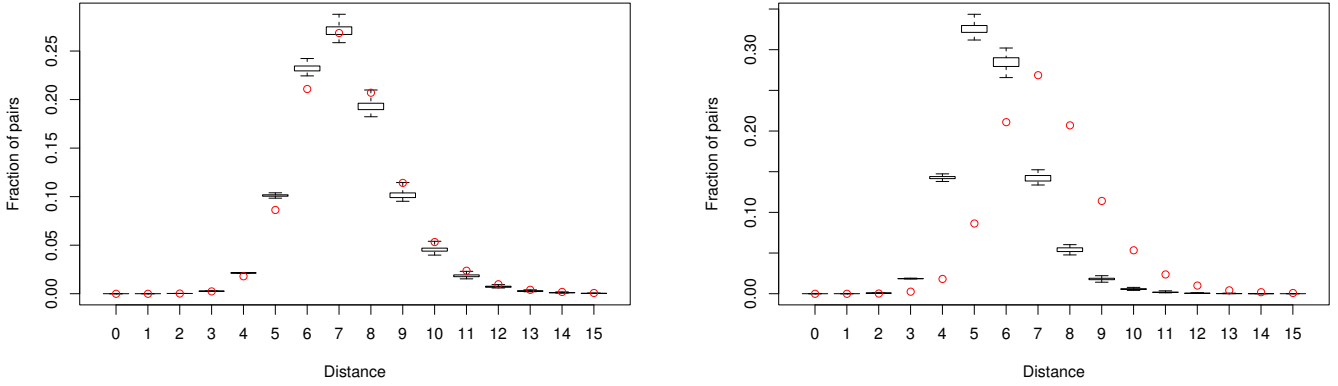

Figure 2: The distribution of pairwise distances $S_{\text{PDD}}$; the small (red) dots correspond to the distribution in the real dblp graph; the boxplots give the distributions for the case $k = 20$, $\varepsilon = 10^{-3}$ (left) and $k = 100$, $\varepsilon = 10^{-4}$ (right). As usual, the two whiskers represent the smallest and largest values observed across the samples, whereas the box represents the range between the lower and the upper quartiles.)

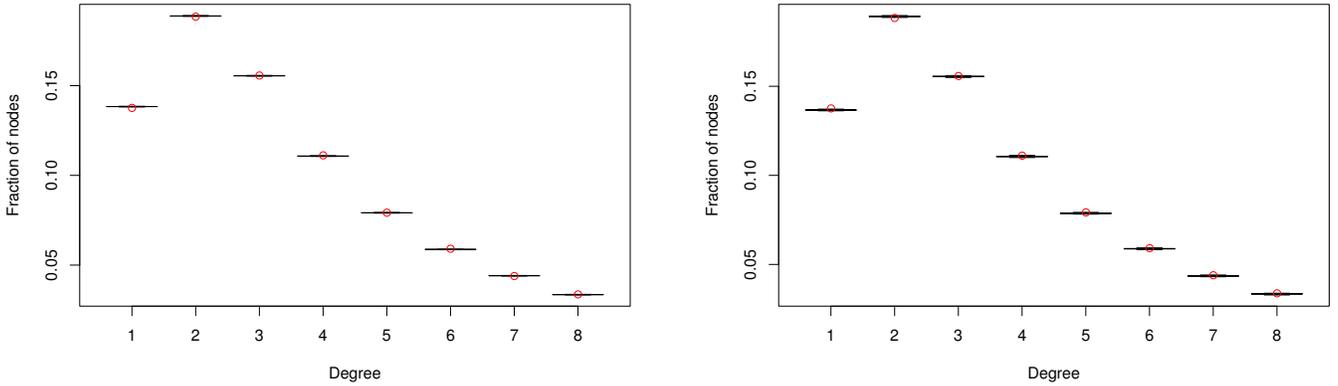

Figure 3: The distribution of degrees $S_{\text{DD}}$; the small (red) dots correspond to the distribution in the real dblp graph; the boxplots give the distributions for the case $k = 20$, $\varepsilon = 10^{-3}$ (left) and $k = 100$, $\varepsilon = 10^{-4}$ (right).

The behavior described for scalar statistics is also observed with vector statistics. For example, Figure 2 shows $S_{\text{PDD}}$ (the distribution of the pairwise distances) in the original dblp and in two obfuscated versions. Here, two extreme cases are presented: For $k = 20$ and $\varepsilon = 10^{-3}$ the distribution obtained is qualitatively very similar (as witnessed also by the scalar distance-based statistics in Table 4); conversely, for $k = 100$ and $\varepsilon = 10^{-4}$, the estimated distribution is quite far from the original one. In Figure 3 we present a similar plot for the degree distribution: for every degree, we considered the distribution of the frequency of that degree across all possible worlds. In this case, the approximation is very concentrated and its mean almost coincides with the real degree frequency, even for $k = 100$ and $\varepsilon = 10^{-4}$.

### 7.3 Comparative Evaluation

We finally compare our proposed method with random-perturbation methods that publish a standard graph (in articular the methods described by Bonchi et al. [4]):

- *random sparsification:* given a parameter $p$, each edge $e \in E$ is removed from the graph with probability $p$;
- *random perturbation:* given a parameter $p$, first each edge $e \in E$ is removed from the graph with probability $p$, then each non-existing edge in $V_2 \setminus E$ is added with probability $\frac{p|E|}{\binom{|V|}{2} - |E|}$ .

To make the comparison possible, we must first determine which value of the parameter $p$ used in these obfuscation algorithms corresponds to which pair $(k, \varepsilon)$ of obfuscation parameters. The appropriate values can be deduced by the anonymity level plots of the sparsified or perturbed graph obtained with a certain value of $p$: of course, any such graph will correspond to many pairs of parameters $(k, \varepsilon)$; for example, given any fixed $\varepsilon$, an appropriate $k$ can be determined by disregarding the $\varepsilon n$ vertices with smallest anonymity and letting $k$ be the least anonymity of the remaining vertices.

Figure 4 shows the obfuscation levels obtained for some of the parameter combinations on dblp and flickr. The plot shows, for every obfuscation level $k$, the number of vertices that have obfuscation level less than or equal to $k$. The two rectangles appearing in the plot highlight the obfuscation requirements $(k, \varepsilon)$. Figure 4 shows, for example, that a random perturbation of `dblp` with $p = 0.04$ matches obfuscation ($k = 60, \varepsilon = 10^{-3}$).

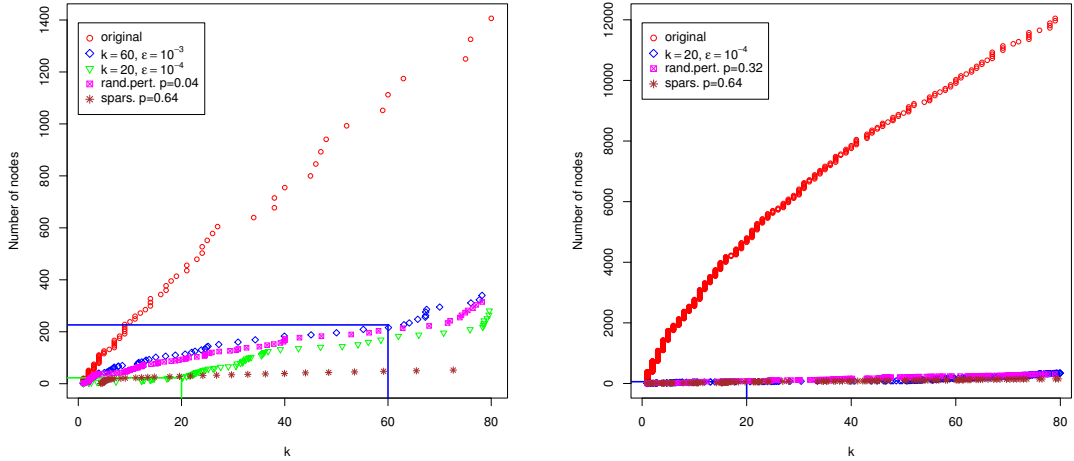

**Figure 4:** Comparison of the anonymity levels obtained for `dblp` (left) and `flickr` (right) using obfuscation, random perturbation and sparsification, for the parameter choices described in Section 7.3. The plot shows, for every obfuscation level $k$, the number of vertices that have obfuscation level less than or equal to $k$.

**Table 6:** Comparison between obfuscation by uncertainty and obfuscation by random sparsification and perturbation.

| | graph | $S_{\text{NE}}$ | $S_{\text{AD}}$ | $S_{\text{MD}}$ | $S_{\text{DV}}$ | $S_{\text{PL}}$ | $S_{\text{APD}}$ | $S_{\text{DiamLB}}$ | $S_{\text{EDiam}}$ | $S_{\text{CL}}$ | $S_{\text{CC}}$ | rel. err. |
|---|---|---|---|---|---|---|---|---|---|---|---|---|
| dblp | original | 716 460 | 6.33 | 238 | 76.79 | $-0.046$ | 7.34 | 25 | 8.78 | 6.96 | 0.38 | |
| | rand.pert. ($p = 0.04$) | 716 393 | 6.33 | 230 | 71.26 | $-0.048$ | 7.09 | 18.55 | 7.25 | 6.85 | 0.36 | 0.071 |
| | obf. ($k = 60, \varepsilon = 10^{-3}$) | 713 819 | 6.31 | 236 | 75.86 | $-0.046$ | 7.15 | 22.75 | 7.21 | 6.82 | 0.36 | 0.043 |
| | rand.spars. ($p = 0.64$) | 257 890 | 2.28 | 93 | 11.40 | $-0.124$ | 10.24 | 36.72 | 10.60 | 25.77 | 0.13 | 0.921 |
| | obf. ($k = 20, \varepsilon = 10^{-4}$) | 713 952 | 6.31 | 233 | 76.18 | $-0.046$ | 7.01 | 22.59 | 7.16 | 6.68 | 0.35 | 0.050 |
| flickr | original | 5 801 442 | 19.73 | 6 660 | 6 200 | $-0.002$ | 5.03 | 21 | 5.43 | 4.80 | 0.12 | |
| | rand.pert. ($p = 0.64$) | 5 801 229 | 19.73 | 2 407 | 820.3 | $-0.0059$ | 4.55 | 7.02 | 4.15 | 4.49 | 0.030 | 0.497 |
| | rand.spars. ($p = 0.32$) | 3 944 902 | 13.41 | 4 526 | 2 871 | $-0.003$ | 5.24 | 19.56 | 4.91 | 6.69 | 0.079 | 0.286 |
| | obf. ($k = 20, \varepsilon = 10^{-4}$) | 5 921 470 | 20.14 | 5 847 | 6 924 | $-0.002$ | 4.84 | 20.51 | 4.81 | 4.64 | 0.050 | 0.112 |

We here present the comparative results in the following cases:[6]

- `dblp` with random perturbation using $p = 0.04$, matching $k = 60$ and $\varepsilon \approx 10^{-3}$;
- `dblp` with sparsification using $p = 0.64$, matching $k = 20$ and $\varepsilon \approx 10^{-4}$;
- `flickr` with random perturbation using $p = 0.32$ and with sparsification using $p = 0.64$, both corresponding to $k = 20$ with $\varepsilon \approx 10^{-4}$.

For each of the two obfuscation techniques presented in [4], we produced 50 samples; note that in those probabilistic methods, the obfuscation is a certain graph. Then we computed the statistics on each sample, and proceeded in the same way as we did for the obfuscated graph.

Table 6 shows the results of the comparison. In all cases, the quality of the statistics as computed with our obfuscation method is much better; in one case, the relative error is 5% instead of the 92% imposed by sparsification to obtain the same level of obfuscation. Therefore, we can safely conclude that our experimental assessment on real-world graphs confirms the initial and driving intuition underlying this paper: by using finer-grained perturbation operations, such as only perturbing *partially* the existence of an edge, one can achieve the same desired level of obfuscation with smaller changes in the data than when completely removing or adding edges, thus maintaining higher data utility.

---

[6]The values of $p$ used here ($p \in \{0.04, 0.32, 0.64\}$) are the same as those used by Bonchi et al. [4].

## 8. CONCLUSIONS AND FUTURE WORK

We introduce a new approach for identity obfuscation in graph data. In the proposed approach, the desired obfuscation is obtained by injecting uncertainty in the social graph and publishing the resulting uncertain graph. Our proposal can be seen as a generalization of random perturbation methods for identity obfuscation in graphs, as it enables finer-grained perturbations than fully removing or fully adding edges. Such increased flexibility in spreading the noise over the edges of the graph enables achieving the same level of obfuscation with smaller changes in the data, as confirmed by our experiments on real-world graphs.

While the results that we achieve are most encouraging, this work represents only a first step in a promising research direction. As it is often the case, new privacy-enabling techniques create novel attacks that, in turn, propel stronger protection mechanisms. Therefore, in our future investigation we plan to extend and strengthen this line of research by further assessing its limits and merits.

One interesting research direction is to investigate how to extend our uncertainty-based approach in order to release networks with additional information, besides the mere graph data, such as vertex attributes [22], communication logs among users, information-propagation traces, and other types of social dynamics. Another case of particular interest is that of a sequential release of a social network. In a recent paper, Medforth and Wang [21] demonstrated the risks of publishing a sequence of releases of the same network. In particular, they described the *degree-trail attack*, by which the vertex belonging to a target user can be re-identified from a sequence of published graphs, by comparing the degrees of the vertices in the published graphs with the degree evolution of the target. The applicability of the degree-trail attack to our probabilistic graph release is an open research question.

**Acknowledgments.** This research was partially supported by the Torres Quevedo Program of the Spanish Ministry of Science and Innovation, co-funded by the European Social Fund, and by the Spanish Centre for the Development of Industrial Technology under the CENIT program, project CEN-20101037, "Social Media" (http://www.cenitsocialmedia.es/). Part of the work was done while P. Boldi and T. Tassa were visiting Yahoo! Research.